\documentclass[aps,prb,floatfix,epsfig,twocolumn,preprintnumbers]{revtex4}
\usepackage{amsfonts}
\usepackage{amssymb}
\usepackage{graphicx}
\usepackage{amsmath}
\usepackage{tablefootnote}
\usepackage{setspace}

\begin{document}

\title{Vertex corrections in self-consistent GW$\Gamma$ calculations: ground state properties of vanadium}
\author{Andrey L. Kutepov\footnote{e-mail: akutepov@bnl.gov}}
\affiliation{Condensed Matter Physics and Materials Science Department, Brookhaven National Laboratory, Upton, NY 11973-5000, USA}

\begin{abstract}
Self-consistent vertex corrected GW calculations have been performed to evaluate the equilibrium volume and corresponding bulk modulus of 3d transition metal vanadium. The study demonstrates the feasibility of this approach. The accuracy of the results obtained for vanadium is considerably better as compared to the accuracy of the results obtained with self consistent GW method without vertex corrections. It is shown that vertex corrections are important not only for the self energy, but also for the irreducible polarizability. The contribution to the value of the $\Psi$-functional from the vertex corrections is about 30 times smaller than the contribution from the GW part (correlation only!), suggesting the fast convergence of the expansion in terms of the screened interaction for this material. Consideration of the Ward-Takahashi Identity shows that the first order vertex correction eliminates considerable (about 80\%) part of the mismatch found in scGW calculations. The remaining small mismatch in the Ward-Takahashi Identity leaves, however, a possibility of the calculated low energy electronic structure being affected by higher order diagrams (not included in this study).
\end{abstract}

\maketitle

\section*{Introduction}
\label{intro}

Below, a brief account is given of the study which purpose was to fill in a certain gap still existing in the applications of the many body perturbation theory (MBPT) to the calculations of the physical properties of materials in their solid state. Essentially, the story is the following. When one begins to study a certain material computationally, it is quite natural to apply first the density functional theory (DFT\cite{pr_136_B864,pr_140_1133}) in its local density approximation (LDA\cite{prl_45_566,cjp_58_1200}) or in the generalized gradient approximation (GGA\cite{prb_33_8800,prl_77_3865}). The approach is rather inexpensive and provides quite reasonable answers to many questions. It is also quite natural and already is becoming a tradition to call for the MBPT when some features of one-electron spectra (for instance, calculated band gap in a semiconductor) are not good enough. As a rule, one applies the simplest MBPT variant, namely the GW approximation (G is for the Green's function, and W is for the screened interaction) in its non-self-consistent form G$_{0}$W$_{0}$.\cite{pr_139_A796,prb_34_5390} If we discuss the band gaps in simple semiconductors/insulators then this approximation allows one to reduce the DFT error considerably (from about 100\% to 10-20\%\cite{prb_93_115203}). Further improvement possible with application of self-consistent vertex corrected GW approach.\cite{prb_95_195120} However, the situation is quite different when one encounters the problem in an application of DFT to the study of the ground state properties (GSP), such as volume dependence of the total energy. There is no well known route yet to cure such a problem. In such situation, the common way is to search for another approximation to the density functional, which can be a meta-GGA\cite{prl_91_146401,prl_103_026403,jcp_125_194101} (addition of the kinetic energy density to the functional) or different flavors of the hybrid functionals\cite{jcp_110_6158,jcp_98_5648,jcp_125_224106,jctc_9_1631} (addition of a portion of non-local exchange). Whereas one can often find an appropriate functional (for a specific material!), the approach hardly can have a predictive power as it looks more like an adjustment to the known experimental results.\cite{jctc_10_3832} Another popular approach in problematic situations is to invoke such methods as DFT+U\cite{prb_48_16929} or DFT+DMFT,\cite{rmp_68_13} and to use on-site Hubbard parameters U and J to improve DFT results for the GSP. However, one has to classify these two methods as semiempirical ones. Often used for the evaluation of U constrained Random Phase Approximation\cite{prb_70_195104} (cRPA) has, as an independent method, its own parameters which can alter the colculated U value by more than factor of two.\cite{prb_82_045105} Whereas a new physics can be brought into consideration in these two approaches (especially in DFT+DMFT), the uncertainty of the on-site (local) approximation and the arbitrariness in the choice of the Hubbard parameters cannot be thought of as totally satisfactory. Namely, often it is not clear which part of the improvement (as compared to DFT) comes from the new physics, and which part comes from the adjustment of the parameters. 

Ab-initio many body methods (those, without involvement of the adjustable parameters and not relying on the on-site (local) approximation) can, in principle, be used to study the GSP. Formal theory for such applications was developed long time ago.\cite{pr_118_1417,AGD} Obviously, principal obstacle up to a present day, was the high computational requirements for applications of ab-initio MBPT to GSP studies in real materials. We are witnessing, however, a great improvement in the computational capabilities nowadays, which can potentially make the applications of MBPT to the studies of GSP possible. Thus, at least from the academic point of view (and from the point of view of possible practical applications in a near future) it is interesting to see how the ab-inito MBPT performs on realistic materials. Existing applications of MBPT for studying the GSP of crystalline materials are not yet numerous. Most of them were non self consistent and used DFT one-electron energies and wave functions as input to evaluate the total energy in Random Phase Approximation\cite{prb_66_245103,prl_103_056401,prl_105_196401,prb_87_214102} (RPA) to the Luttinger-Ward functional.\cite{pr_118_1417} A few years ago, fully self consistent GW approach was applied to study the GSP of simple elements (Na, Al, Si) in their crystalline state.\cite{prb_80_041103}

Recently, self-consistent GW calculations (scGW) were performed for 3d transition metals.\cite{jcm_29_465503} scGW approach represents the simplest variant of ab-inito self-consistent MBPT suitable for studying GSP. The calculations have demonstrated improved performance as compared to the calculations in LDA. The error in the calculated equilibrium volume was reduced by a factor of more than 2 as compared to the LDA error. But the performance of scGW method in comparison with the performance of GGA was not very good. However, quite encouraging circumstance found in [\onlinecite{jcm_29_465503}] is that the deviation (for example of the calculated equilibrium volume) from the experimental data is quite systematic in scGW and it is suggestive of a certain well defined effect which is missing in scGW approximation. It was speculated in [\onlinecite{jcm_29_465503}] that vertex corrections might be a good candidate for this missing effect. The speculation was based on the observation\cite{prb_96_035108} that in the electron gas, the gain in the calculated total energy when vertex corrections are included becomes more negative when the volume expands. This work is designated for a systematic study of this question with the metal Vanadium as an example. The reason why this element has been picked up from the 3d metals is that it is the simplest (one atom per unit cell and no magnetic order) which is crucial for very time consuming self-consistent vertex corrected (scGW$\Gamma$) calculations.

The plan of the paper is the following. Section \ref{methods} explains the method which was used in the study and the computational setup. Section \ref{res} provides the results obtained and the discussion. The conclusions are given afterwords.

\section{Calculation method and setup}
\label{methods}

When generating conserving self consistent schemes which include vertex corrections, it is convenient to start with the variational expression for the grand potential written in terms of the so called $\Psi$-functional introduced by Almbladh et al. in [\onlinecite{ijmpb_13_535}]:

\begin{align}\label{om0}
\Omega=\Omega_{0}+\Psi&+\frac{1}{2}Tr[PW+\ln(1-VP)]\nonumber\\&-Tr[G^{-1}_{0}G-1]+Tr[\ln G-\ln G_{0}],
\end{align}
where $G_{0}$ is the Green function of some reference (non-interacting) system, and $\Omega_{0}$ is the corresponding grand potential. The $\Psi$-functional is a functional of the Green's function G and the screened interaction W, and is related to the Luttinger-Ward $\Phi$-functional\cite{pr_118_1417}:

\begin{equation}\label{om1a}
\Psi [G,W]=\Phi[G]-\frac{1}{2}[PW-\ln (1+PW)],
\end{equation}
where $P$ is the irreducible polarizability. The advantage of using the $\Psi$-functional instead of the Luttinger-Ward functional in the approximations which include vertices is that the expansion in terms of W converges much faster than the expansion in terms of the bare Coulomb interaction V. As a result, a number of useful approximations with non-trivial vertices can be formulated in terms of just a few diagrams. On the other hand, in the $\Phi$-functional formulation which uses expansion in terms of the bare Coulomb interaction, even the GW approximation which corresponds to the trivial vertex, consists of an infinite number of diagrams (ring diagrams). Taking an approximation to the $\Psi$-functional (by drawing a set of the diagrams with G and W as building blocks) the approximations for the irreducible polarizability and the self energy $\Sigma$ are fixed by the relations\cite{ijmpb_13_535}:

\begin{equation}\label{om1b}
P=-2\frac{\delta \Psi}{\delta W}|_{G},
\end{equation}
and
\begin{equation}\label{om1c}
\Sigma=\frac{\delta \Psi}{\delta G}|_{W}.
\end{equation}

In this study, the Hartree-Fock (exchange) approximation was used as a reference and, correspondingly, the expression for the grand potential was the following:

\begin{align}\label{om1}
\Omega=\Omega_{x}+\Psi_{c}&+\frac{1}{2}P\widetilde{W}+\frac{1}{2}Tr[PV+\ln(1-VP)]
\nonumber\\&-Tr[G^{-1}_{x}G-1]+Tr[\ln G-\ln G_{x}],
\end{align}

where the exchange part (first term on the right hand side (rhs)) and the correlation part (the remaining terms on the rhs) were separated for convenience, and $G_{x}$ is the exchange part of Green's function. If we further represent the full irreducible polarizability as a sum of a simple bubble ($P_{0}$) and the vertex part $\Delta P$: $P=P_{0}+\Delta P$, then the very first diagram in $\Psi_{c}$ (which is $-\frac{1}{2}TrP_{0}\widetilde{W}$, with $\widetilde{W}$ being the frequency-dependent part of the $W$: $W=V+\widetilde{W}$) can be canceled with the corresponding contribution from the third term on the rhs of Eq.(\ref{om1}). Equation (\ref{om1}) can, therefore, be written in more convenient for the evaluation form:

\begin{align}\label{om11}
\Omega=\Omega_{x}+\Psi_{vrt}+&\frac{1}{2}\Delta P\widetilde{W}+\frac{1}{2}Tr[PV+\ln(1-VP)]
\nonumber\\&-Tr[G^{-1}_{x}G-1]+Tr[\ln G-\ln G_{x}],
\end{align}
where $\Psi_{vrt}$ stands for the part of $\Psi$-functional containing non-trivial vertices. The exchange part of the grand potential entering the equation (\ref{om11}) is:

\begin{align}\label{om2}
\Omega_{x}=&-\frac{1}{\beta N_{\mathbf{k}}}\sum_{\alpha \mathbf{k}\lambda} \ln 
[1+e^{-(\varepsilon^{\alpha\mathbf{k}}_{\lambda}-\mu)\beta}]+E_{coul}-Tr (V^{H}G)\nonumber\\
&-\frac{1}{2} Tr (\Sigma^{x}G),
\end{align}
where $\alpha$ is the spin, $\mathbf{k}$ enumerates k-points in the Brillouin zone, $\lambda$ is the band index, $\beta$ is the inverse temperature, $N_{\mathbf{k}}$ is the total number of k-points in the Brillouin zone, $\mu$ is the chemical potential. $E_{coul}$ represents the energy of the Coulomb interaction (sum of nuclear-nuclear, electron-nuclear, and classical electron-electron). $V^{H}$ is the electronic Hartree potential, $\Sigma^{x}$ is the exchange (frequency-independent) part of self energy.

\begin{figure}[t]
\centering
\includegraphics[width=7.0 cm]{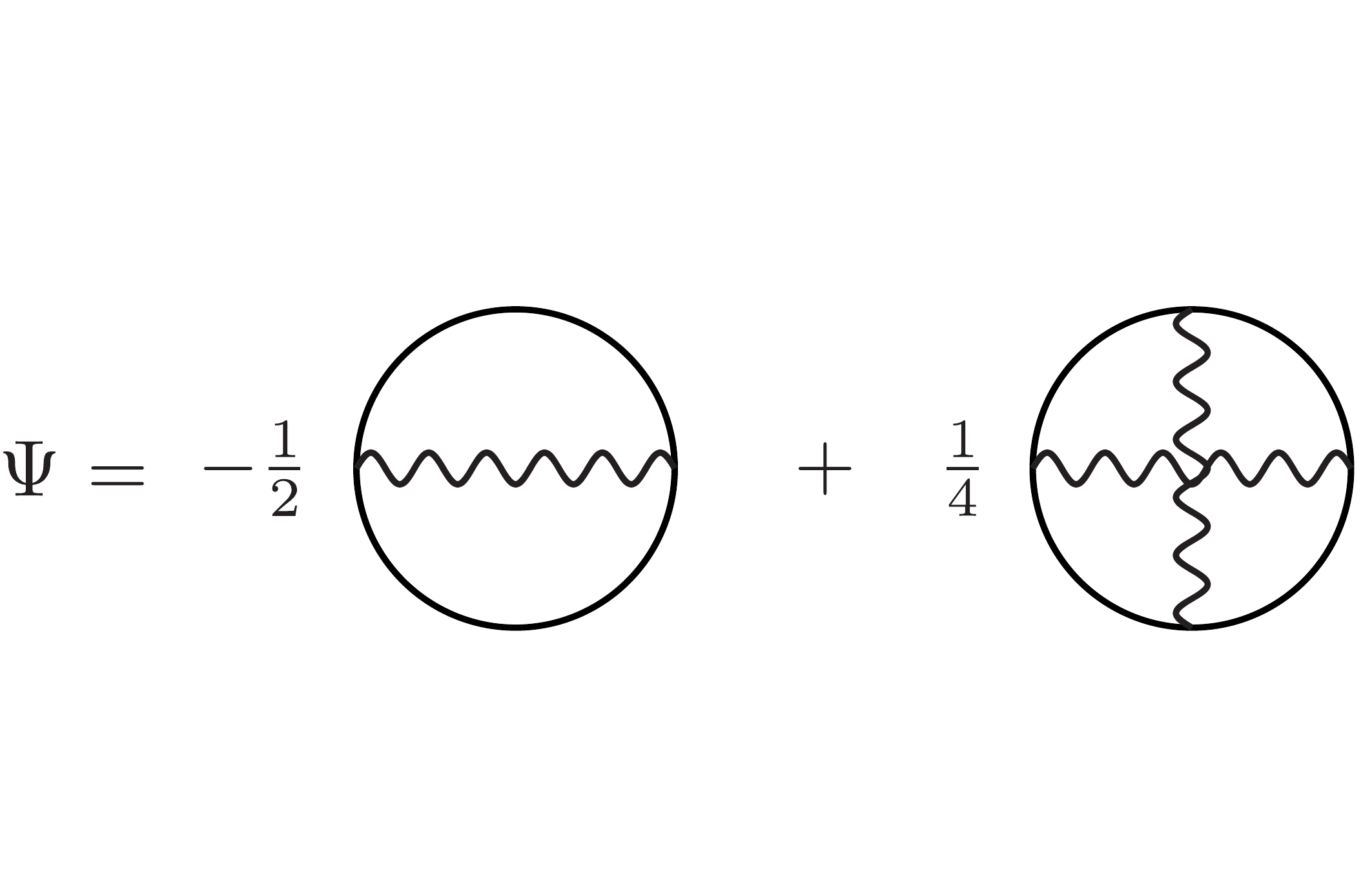}
\caption{Scheme of the evaluation of the second order diagram for the self energy.}
\label{diag_Psi}
\end{figure}

In this work, in order to generate an approximation for the correlation part of the problem, the simplest approximation for the $\Psi$-functional which includes vertex correction has been adapted (Fig. \ref{diag_Psi}). In Fig. \ref{diag_Psi}, the first diagram corresponds to the famous GW approximation, whereas the second one includes first order vertex correction. The second diagram in Fig. \ref{diag_Psi} is, correspondingly, the quantity $\Psi_{vrt}$ which was introduced before in Eq. (\ref{om11}).

\begin{figure}[b]
\centering
\includegraphics[width=7.0 cm]{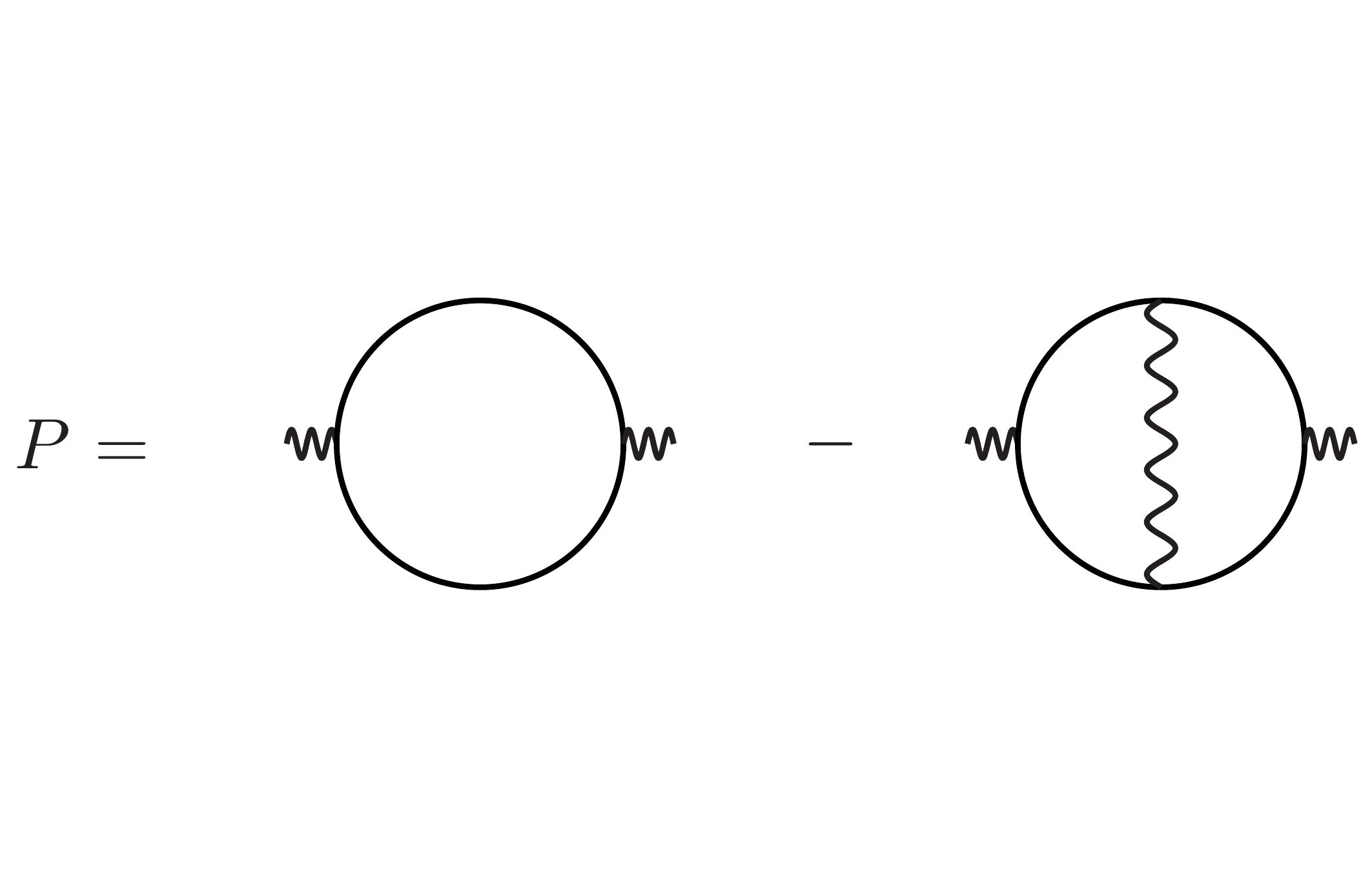}
\caption{Diagrammatic representation of the irreducible polarizability in the simplest vertex corrected scheme.}
\label{diag_P}
\end{figure}

\begin{figure}[t]
\centering
\includegraphics[width=7.0 cm]{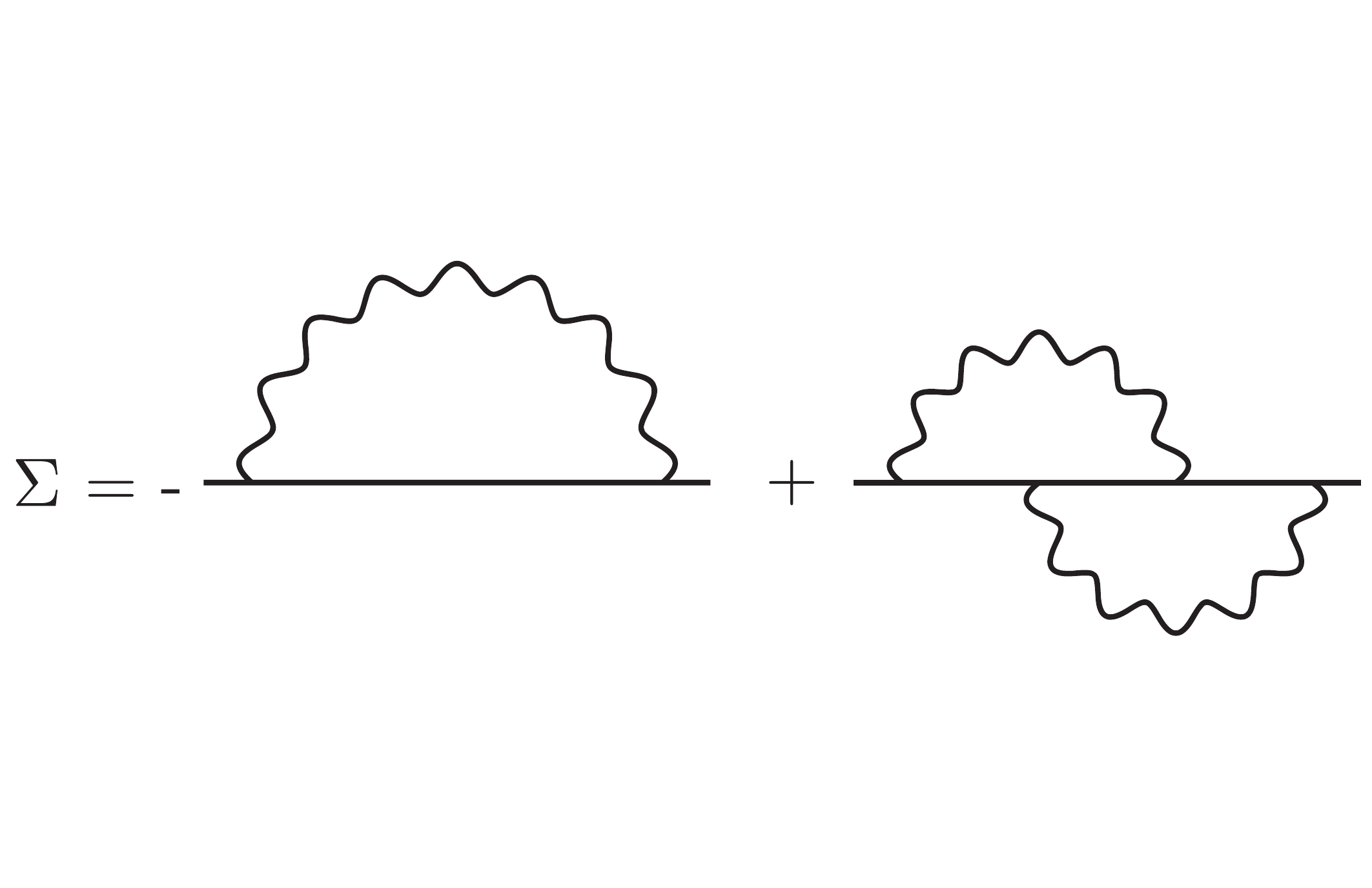}
\caption{Diagrammatic representation of the self energy in the simplest vertex corrected scheme.}
\label{diag_S}
\end{figure}

Diagrammatic representations for the irreducible polarizability (Fig. \ref{diag_P}) and for the self energy (Fig. \ref{diag_S}) follow from the chosen approximation to the $\Psi$-functional. The set of diagrams for the polarizability and the self energy shown in Figs. \ref{diag_P} and \ref{diag_S} corresponds to the scheme B introduced earlier in Ref. [\onlinecite{prb_94_155101}]. Technical details of the GW part were described in the Ref. [\onlinecite{prb_85_155129}]. Numerical algorithm for the evaluation of the first order polarizability was the same in this study as described in details in Ref. [\onlinecite{prb_94_155101}]. For the evaluation of the second order self-energy, however, more efficient algorithm (as compared to the one described in [\onlinecite{prb_94_155101}]) was used. The brief account of the details of this new algorithm can be found in Appendix. The value of $\Psi_{vrt}$ can be evaluated from the second order self energy $\Sigma_{2}$ (second diagram in Fig. \ref{diag_S}) as its convolution with Green's function: $\Psi_{vrt}=\frac{1}{4}Tr(\Sigma_{2}G)$. Equivalently, the $\Psi_{vrt}$ can be evaluated from the first order irreducible polarizability $P_{1}$ (second diagram in Fig. \ref{diag_P}) as its convolution with the screened interaction: $\Psi_{vrt}=-\frac{1}{2}Tr(P_{1}W)$.

For the remaining terms in Eq.(\ref{om11}), the following expressions were used in practical calculations:

\begin{equation}\label{om4}
Tr[G^{-1}_{x}G-1]=Tr[\Sigma_{c}G],
\end{equation}
and
\begin{equation}\label{om5}
Tr[\ln G-\ln G_{x}]=Tr_{0} \sum_{\omega>0}\ln [G(\omega)G^{+}(\omega)-G_{x}(\omega)G^{+}_{x}(\omega)].
\end{equation}

In the equations above, $\Sigma_{c}$ is the correlation (frequency-dependent) part of the self energy. Matrices in Eq. (\ref{om5}) have been transformed to the hermitian form in order to facilitate the evaluation of the logarithm. Taking the trace ($Tr$) means the summation over the \textbf{k}-points in the Brillouin zone, spins, Matsubara's frequencies, and diagonal matrix elements. Symbol $Tr_{0}$ means that frequency summation is excluded from the evaluation of trace. Having evaluated the grand potential, the free energy simply is $F=\Omega-\mu N$, with $N$ being the number of electrons. In order to obtain the equilibrium volume and the corresponding bulk modulus, free energy was evaluated on the equidistant mesh of six volumes arranged near the theoretical minimum of the free energy. The obtained function $F(V)$ was approximated by quadratic polynomial of $V$ which allowed one to get the position of the minimum (equilibrium volume) and the second derivative (related to the bulk modulus).

Bloch band states of the effective exchange problem\cite{prb_85_155129} (corresponds to the approximation $\Sigma^{c}=0$) were used to represent $G$, $V^{H}$, $\Sigma^{x}$, and $\Sigma^{c}$. The polarizability and the Coulomb interaction were expressed in the mixed product basis.\cite{prb_76_165106,cpc_219_407} FLAPW+LO method was used to solve the one-electron problems (Kohn-Sham equation in LDA and effective exchange hamiltonian). The specifics of the construction of the local orbitals (LO) were given earlier in the Ref. [\onlinecite{jcm_29_465503}]. The number of LO's for every \textit{l} was defined by specifying (from the convergence requirement) the maximal principal quantum number $n_{max}$ which was equal 9 in this study. The maximal value of the orbital momentum in MT spheres was equal 8. As compared to the work [\onlinecite{jcm_29_465503}], the radius of the muffin-tin (MT) spheres was slightly reduced in this study (2.1 a.u. versus 2.47 a.u. in [\onlinecite{jcm_29_465503}]). This was done in order to alleviate the need for using too many LO's inside the MT spheres and, thus, to make the basis set better balanced between the MT spheres and the interstitial region. The number of the augmented plane waves was about 100. Integrations over the Brillouin zone were performed using the mesh $12\times 12\times 12$. However, the results obtained with $8\times 8\times 8$ were only very slightly different (with difference in the calculated equilibrium volume within 0.1\%) Evaluations of the vertex corrections were done on the coarser mesh $4\times 4\times 4$ with subsequent interpolation. All band states generated from the LAPW+LO basis set (about 300) were used to represent the self energy and Green's function of the GW part of the problem. 20 band states closest to the chemical potential were used to evaluate the vertex corrections. The size of the mixed product basis is related to the quality of LAPW+LO basis and was 1040 (average over the \textbf{k}-points) for the GW part and 120 for the vertex part. The electronic temperature $T$ was 500K. 64 imaginary time points (distributed inhomogeneously\cite{prb_85_155129}) were used to represent functions on the interval $[0:1/T]$. Similar number of Matsubara's frequencies (again, distributed inhomogeneously) were used to effectively cover the interval $[-\infty:+\infty]$. One should point out that an infinite number of Matsubara's frequencies was, in fact, used in all frequency sums. The above number (64) just means the number of frequencies used for the interpolation purposes (see the Ref. [\onlinecite{prb_85_155129}] for details). For the selected temperature (500 K), 64 points in the Matsubara's time/frequency domains provide very high level of convergence.

The calculations didn't involve any uncontrollable approximations, such as on-site (local) approximation or static (frequency independent) approximation for W. Besides truncation in the diagrammatic representation of the $\Psi$-functional, the only approximations involved were the ones related to the finite number of \textbf{k}-points, frequency/time points, band states, and product basis functions. All of them are totally controllable. Every calculation included 22 DFT iterations (up to convergence) which purpose was to generate an initial approximation for the Green's function to start scGW iterations. Then 30 iterations of scGW followed which allowed to converge the calculated free energy within $10^{-5} Ry$. Finally, 20 iterations of scGW$\Gamma$ were performed, starting with G and W from the last scGW iteration and providing in the end similar to the scGW convergence in the free energy. 512 MPI processes were used. Approximate time for one scGW iteration was 18 minutes, whereas about 1 hour was required to accomplish one scGW$\Gamma$ iteration. All calculations have been performed using computer code FlapwMBPT.\cite{flapwmbpt}

One should mention here one issue which possibly can be faced when higher order diagrams (beyond GW approximation) are applied. Namely, it was stated\cite{jpc_7_3013, prb_58_12684,prl_80_1702,prb_90_115134} that negative values may appear in the calculated spectral function when using higher order approximations for the self energy.  This issue, however, was not encountered in this study (at least in the range $\pm$20 eV around chemical potential). Nor was it faced in previous studies\cite{prb_94_155101,prb_95_195120} of simple metals and a number of semiconductors/insulators. Our study of the electron gas\cite{prb_96_035108} was also free of this potential problem. Calculated spectral functions were positive also in a few materials (CeO$_{2}$, SrTiO$_{3}$, NiO, TiO$_{2}$, FeSb$_{2}$, unpublished) where second order self energy was used. Whereas there is no guarantee that any material is free of this problem, the issue, thus, doesn't seem to be of a great concern from the practical point of view. However, positivity/negativity of the calculated spectral function can be a good indicator of the accuracy of the numerical methods employed. Namely, it was noticed (not only in scGW$\Gamma$ but also in scGW calculations) that negative spectral functions can actually be obtained when the electronic temperature is too low for a given number of Matsubara's time ($\tau$) points and/or for a given number of \textbf{k}-points in the Brillouin zone. In such cases, simple increase in the number of $\tau$-points or in the number of \textbf{k}-points always restored the positivity of the spectral function.

\section{Results}
\label{res}

\begin{table}[b]
\caption{Theoretical and experimental equilibrium volume and bulk modulus of vanadium. All theoretical results in this Table were obtained with the same code FlapwMBPT. The experimental data corrected for zero-point vibrational effects have been cited from the Ref. [\onlinecite{jctc_10_3832}].} \label{v0b0}
\begin{center}
\begin{tabular}{@{}c c c} &$V_{0}$ (a.u./atom) &$B_{0}$ (GPa)\\
\hline \hline
LDA &84.31 &247.0\\
GGA &92.62 &186.2\\
scGW &88.04 &206.9\\
scGW$\Gamma$ &91.28 &180.1\\
Exp. &91.94 &158.9\\
\hline \hline
\end{tabular}
\end{center}
\end{table}

\begin{figure}[t]
\centering
\includegraphics[width=7.0 cm]{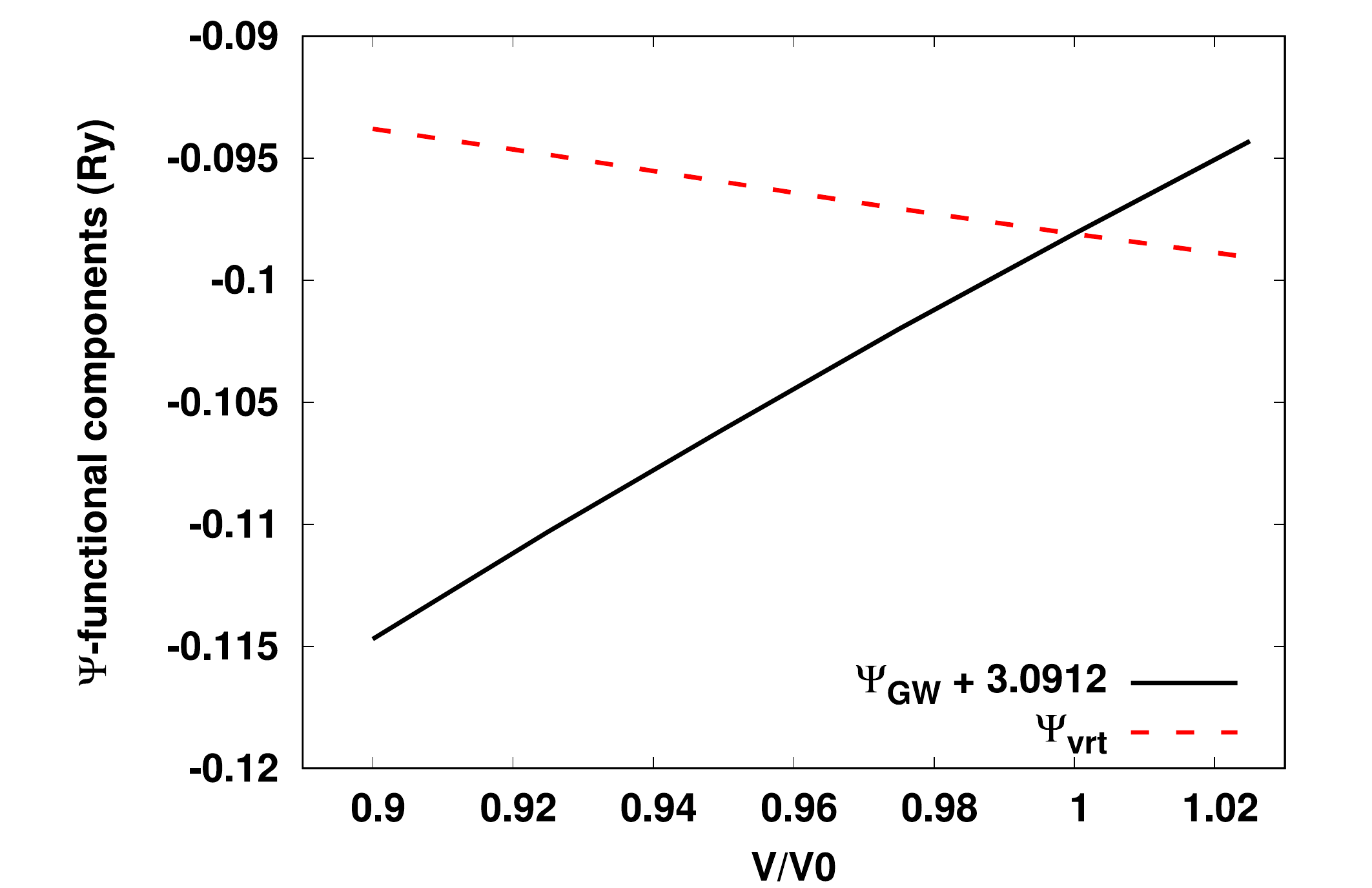}
\caption{Components $\Psi_{GW}$ (GW part, only correlation) and $\Psi_{vrt}$ (vertex part) of the $\Psi$-functional as functions of relative volume for vanadium. The values for the $\Psi_{GW}$ component were shifted upward by 3.0912 Ry for better comparison with the components of the $\Psi_{vrt}$. $V_{0}$ is the experimental equilibrium volume.} \label{psi_vv0}
\end{figure}

\begin{figure}[b]
\centering
\includegraphics[width=7.0 cm]{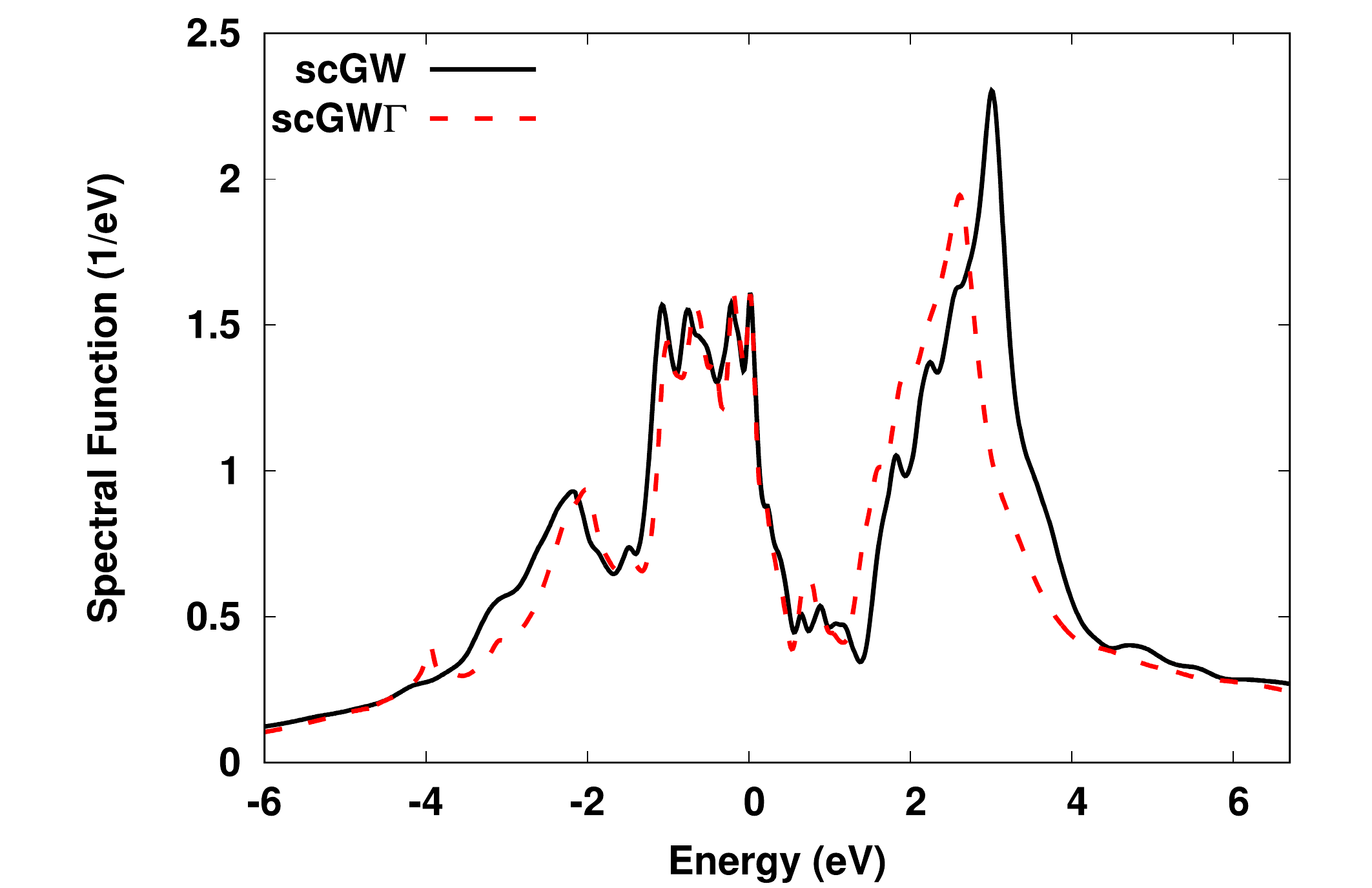}
\caption{Spectral function as obtained in scGW and in scGW$\Gamma$ for the experimental equilibrium volume of Vanadium. Analytical continuation for the frequency dependent part of the self energy was performed as described in the Appendix D of the Ref. [\onlinecite{prb_85_155129}]. It was needed to obtain the spectral function as a function of real frequency.} \label{dos}
\end{figure}

\begin{figure}[b]
\centering
\includegraphics[width=8.0 cm]{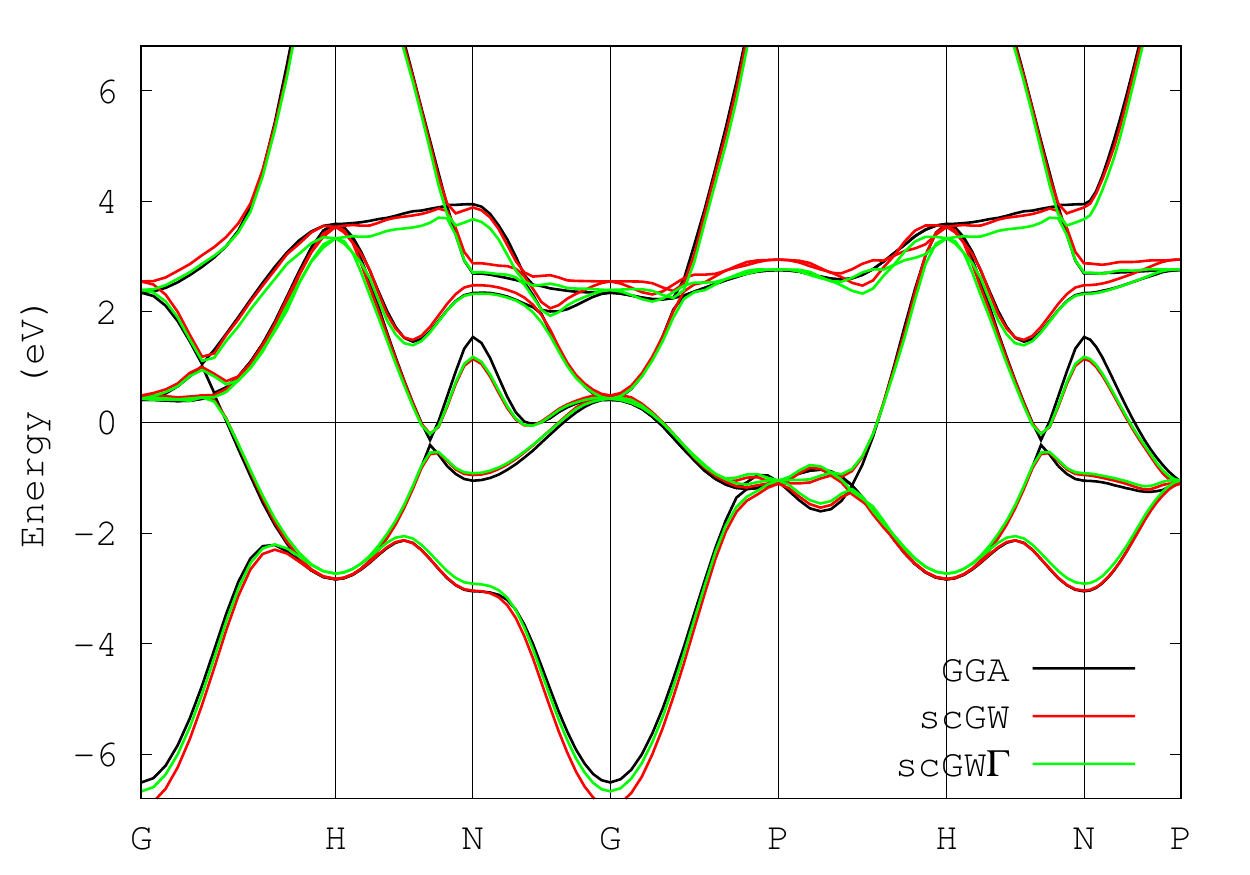}
\caption{Band structure of vanadium as obtained in GGA, scGW, and scGW$\Gamma$ approximations. In case of scGW and scGW$\Gamma$ methods, the data for plotting have been obtained in two steps. As a first step, the effective one-electron energies on the regular mesh of \textbf{k}-points were estimated by a linearization of the self-consistent self energies close to the chemical potential (see discussion in the Appendix D of the Ref.[\onlinecite{prb_85_155129}], where the approach was abbreviated as QP-III). As a second step, Fourier interpolation\cite{pr_178_1419,jcompp_67_253,prb_38_2721} has been used to obtain the band energies along the path in \textbf{k}-space.} \label{bnd}
\end{figure}

Figure \ref{psi_vv0} represents the volume dependence of the GW part (correlation only, i.e. only frequency dependent part of W is counted) and the vertex part of the $\Psi$-functional. Correlation part of the $\Psi_{GW}$ has a positive slope reflecting the fact that the effect of the correlations included in the GW approximation consists primarily in reducing the calculated in the Hartree-Fock approximation (and, as a rule, severely overestimated) equilibrium volume. On the other hand, as one can see from the Fig. \ref{psi_vv0}, the vertex part of the functional has negative slope, and acts in the opposite (as compared to the GW part) direction, increasing the calculated (now in GW approximation) equilibrium volume. Despite the fact, that the amplitude of the $\Psi_{vrt}$ is about 30 times smaller than the amplitude of the correlation part of the $\Psi_{GW}$, its variations around $V_{0}$ are about 5 mRy, which is not negligible and provide an important contribution to the final result for the GSP. Table \ref{v0b0} presents the calculated equilibrium volume and corresponding bulk modulus. The results obtained in LDA and GGA have also been included for comparison. As one can see from the Table, scGW$\Gamma$ eliminates considerable part of the disagreement of scGW and experiment, justifying the speculation about importance of the vertex corrections mentioned in the Introduction. Remaining mismatch with the experimental data is pretty small (about 0.7\% in the equilibrium volume and about 11\% in the bulk modulus). It can be attributed to numerical inaccuracies (especially to the evaluation of the second derivative of the free energy needed for the bulk modulus), and/or to the higher order vertex corrections not included in this study. In any case, the smallness of the mismatch suggests that higher order vertex corrections are not important for the GSP of vanadium, i.e. from this point of view, vanadium belongs to the so called weakly correlated materials. scGW$\Gamma$ results for vanadium, as one can judge, are considerably more accurate as compared to the results obtained in LDA, and have similar accuracy as compared to the GGA results.

Spectral functions presented in Fig. \ref{dos}, as obtained in scGW and in scGW$\Gamma$, do not show any essential difference. The effect of vertex corrections consists in a slight narrowing of the spectral features, namely, the states under the Fermi level are shifted up, and the states above the Fermi level are shifted towards lower energies. The smallness of the effect of vertex corrections on the spectral function supports the idea that vanadium belongs to the class of materials, which can be well addressed with ab-initio MBPT. Similar conclusion can be drawn from the plot of the band structure presented in Fig.\ref{bnd} where only small deviations from scGW bands can be seen in scGW$\Gamma$ results. But one cannot completely ignore a slight renormalization of the bands when one goes from GGA and scGW to scGW$\Gamma$ approach, particularly at the points H and N of the Brillouin zone (up to 10$\%$). Renormalization of the GGA bands is not seen at the $\Gamma$ and P points. In the Ref. [\onlinecite{prb_97_155128}] it was shown, that spin-fluctuations (which were not considered in the present study) can be important for low energy band structure renormalization. However, the approach employed in [\onlinecite{prb_97_155128}] was not self-consistent and it used empirical parameters (Hubbard U). Thus, based on [\onlinecite{prb_97_155128}], one can only make qualitative conclusions about the effect of spin fluctuations. For a quantitative conclusions one should avoid empirical parameters, allowing all interactions in the system to be calculated self-consistently. The investigation of the effect of higher order vertex corrections (including spin fluctuation diagrams) on the band renormalization in 3d and possibly 4d metals, thus, represents an interesting object for future research. But it is beyond the scope of the present work.

\begin{figure}[t]
\centering
\includegraphics[width=7.0 cm]{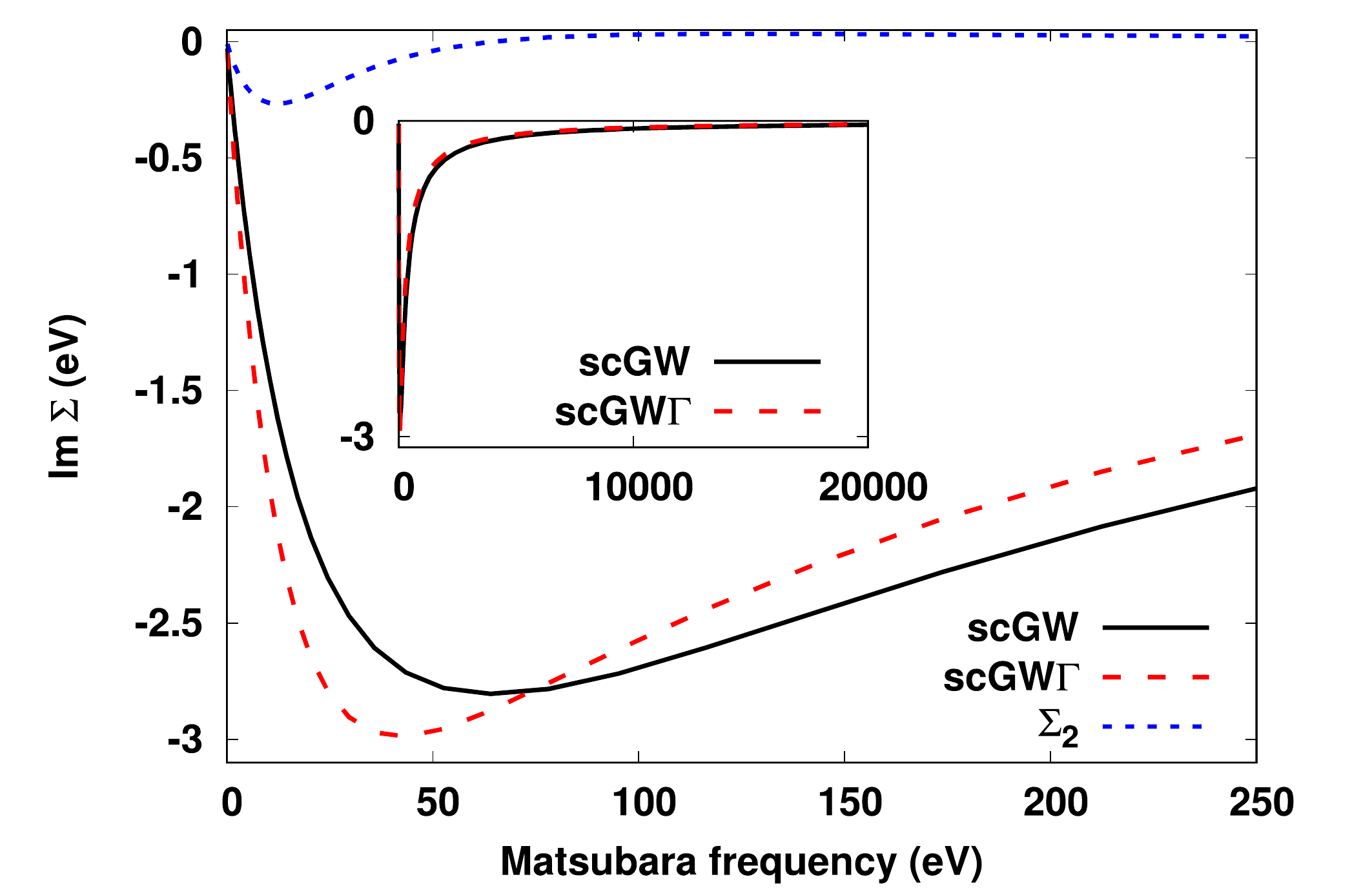}
\caption{Imaginary part of the self energy for the uppermost valence band (s-band) at the $\Gamma$ point in the Brillouin zone. $\Sigma_{2}$ corresponds to the second order contribution to the self energy (last diagram in the Fig. \ref{diag_S}). The inset shows the self energy from scGW and scGW$\Gamma$ calculations in a broader frequency range.} \label{sigma_s}
\end{figure}

\begin{figure}[b]
\centering
\includegraphics[width=7.0 cm]{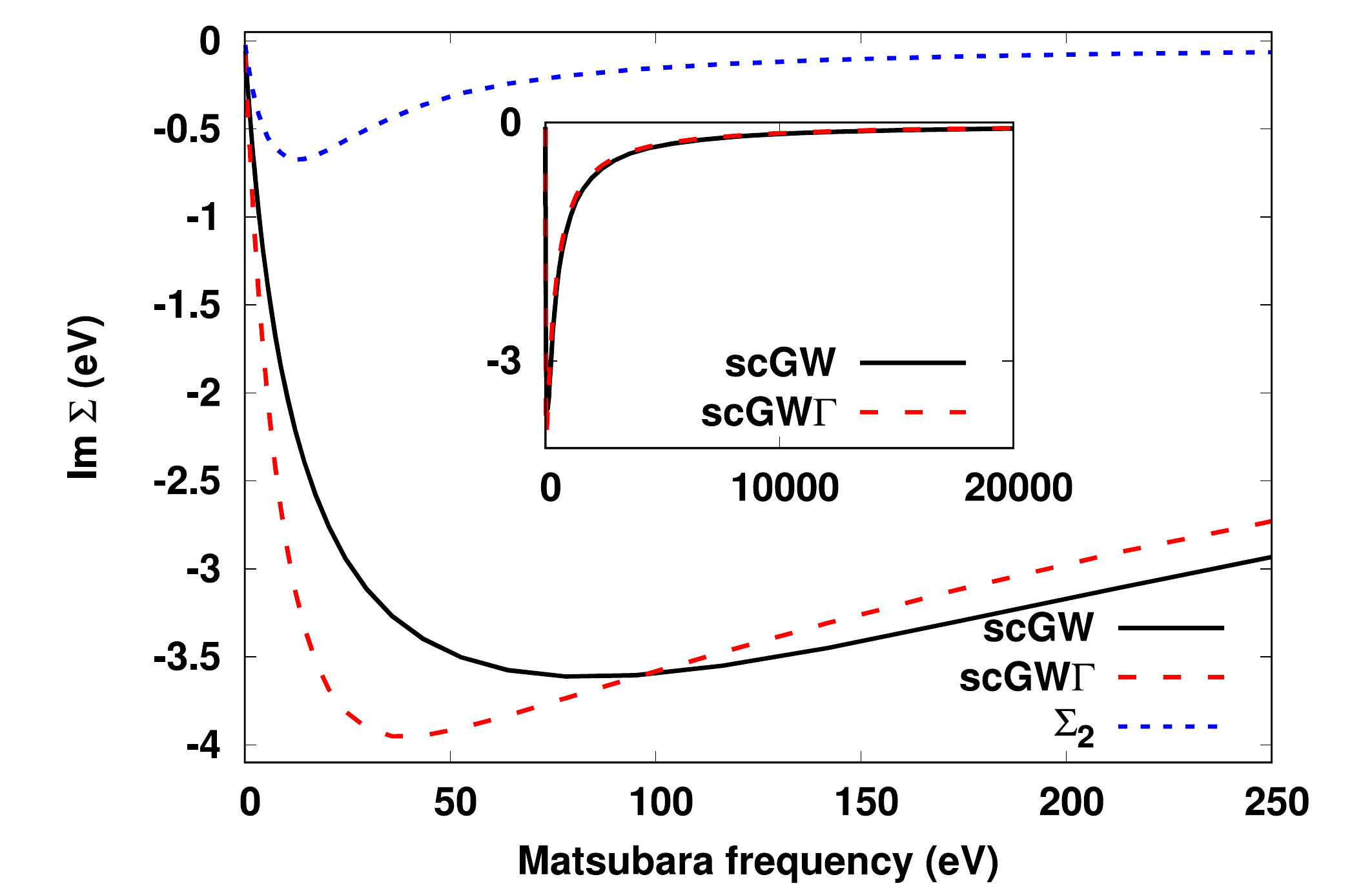}
\caption{Imaginary part of the self energy for the lowest conduction band (d-band) at the $\Gamma$ point in the Brillouin zone. $\Sigma_{2}$ corresponds to the second order contribution to the self energy (last diagram in the Fig. \ref{diag_S}). The inset shows the self energy from scGW and scGW$\Gamma$ calculations in a broader frequency range.} \label{sigma_d}
\end{figure}

\begin{figure}[t]
\centering
\includegraphics[width=7.0 cm]{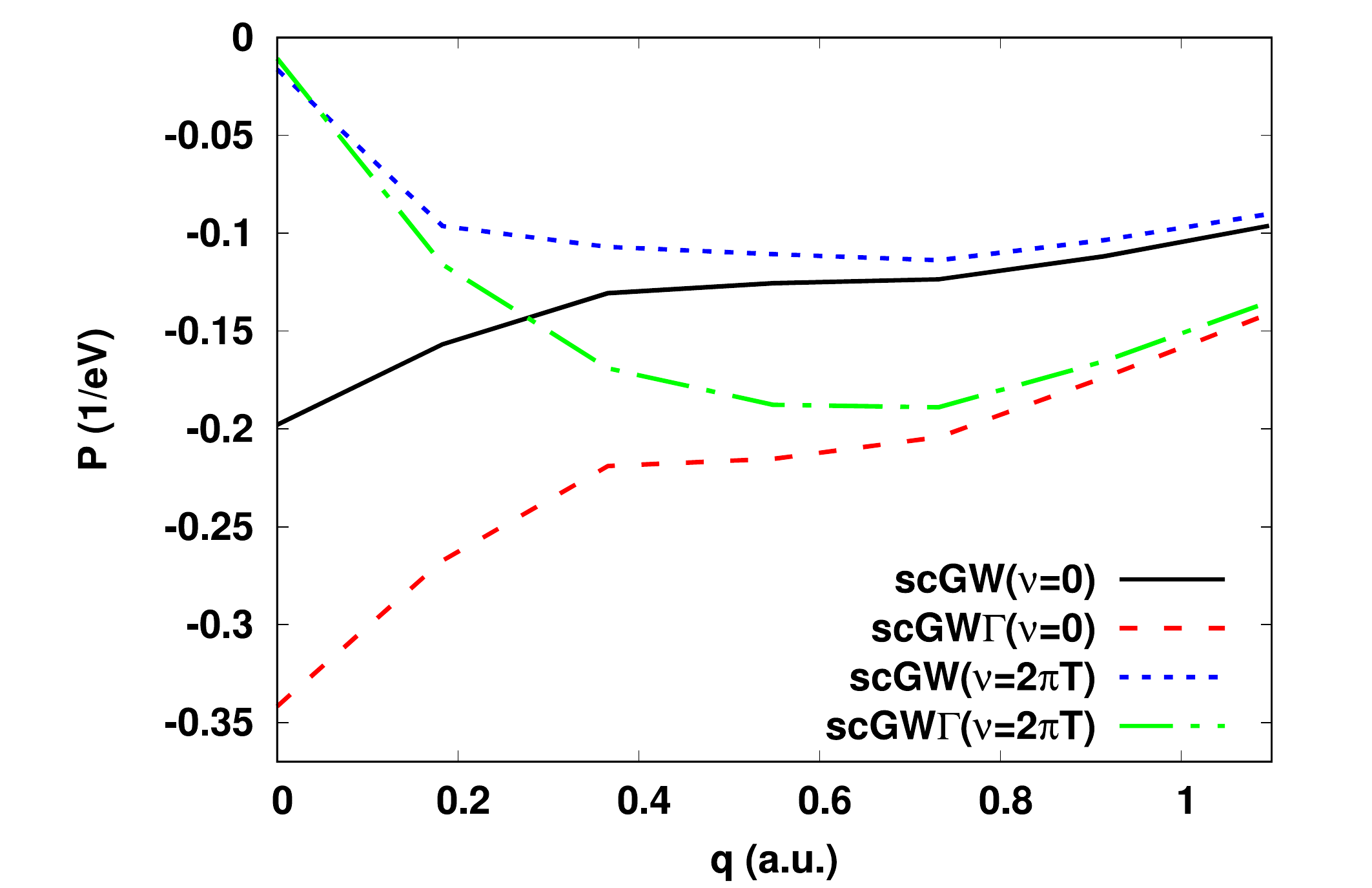}
\caption{The components $P^{\mathbf{q}}_{\mathbf{G}=\mathbf{G}'=0}(\nu)$ of the calculated irreducible polarizability as functions of the momentum $\mathbf{q}$ along the direction (011) in the Brillouin zone.} \label{p_pw}
\end{figure}

\begin{figure}[b]
\centering
\includegraphics[width=7.0 cm]{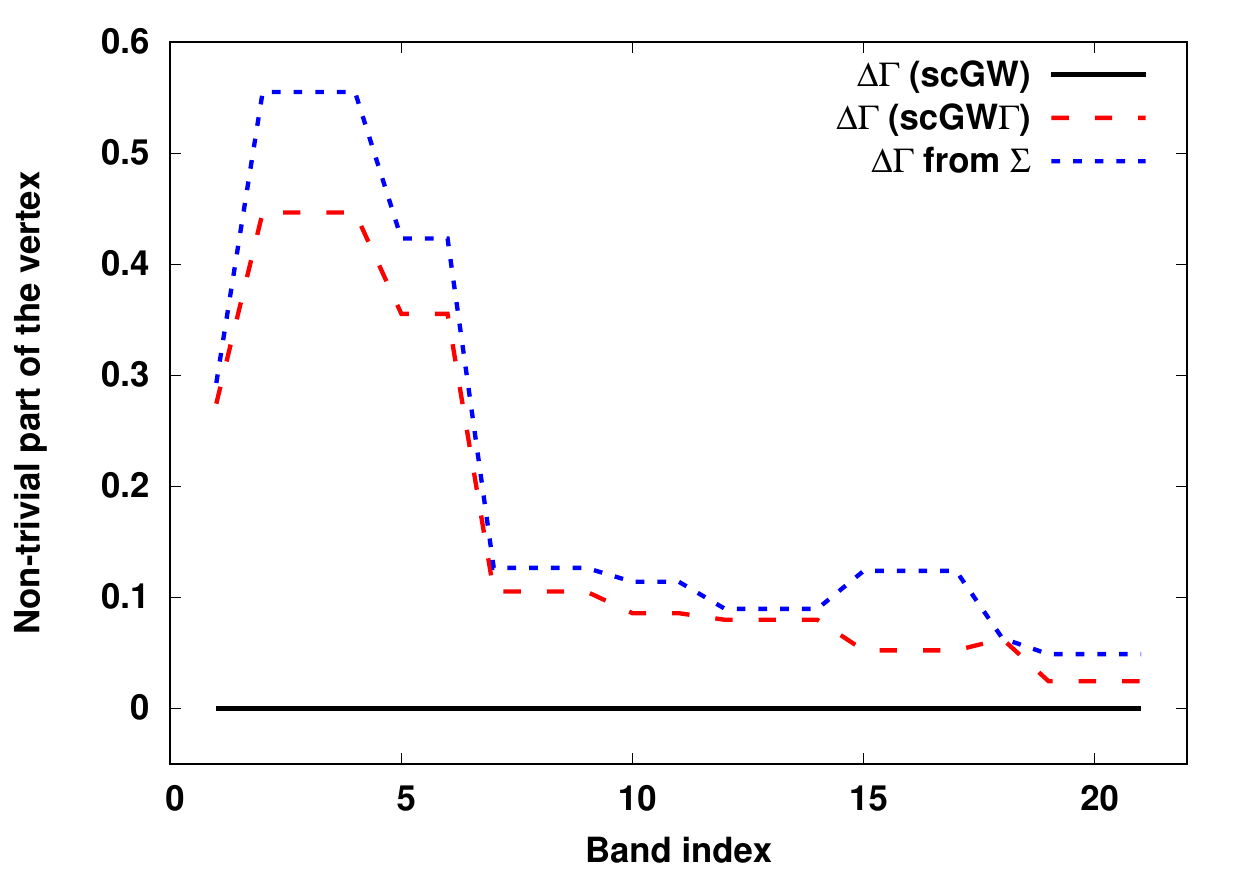}
\caption{Mismatch in the Ward-Takahashi Identity as obtained in scGW and scGW$\Gamma$ approaches. Non-trivial parts of the vertex function (components with $\lambda=\lambda'$ in the Eq. \ref{wti1}) for the 21 bands at the $\Gamma$ point in the Brillouin zone are presented. Counting begins from the lowest valence band. Fermionic Matsubara frequency ($\omega$) was taken to be $\pi T$, whereas the bosonic frequency $\nu$ was $2\pi T$.} \label{wti}
\end{figure}

Figures \ref{sigma_s} and \ref{sigma_d} illustrate the self energy as a function of Matsubara frequency for the uppermost valence band (s-band) and the lowest conduction band (d-band) at the $\Gamma$ point in the Brillouin zone. As one can see, the contribution of the second order diagram essentially disappear above 100 eV, whereas GW part decays very slowly up to a few tens of thousand electron-volts (see insets).

The difference in the self energy obtained from scGW and from scGW$\Gamma$ calculations consists not only from the second order diagram shown in Fig. \ref{diag_S} but also from the vertex correction in W entering through the polarizability and from the effects of self-consistency in both G and W. As it is seen in the figures \ref{sigma_s} and \ref{sigma_d}, the second order self energy correction is effective only for the very low frequencies (roughly up to 50 eV). The noticeable difference in the scGW and scGW$\Gamma$ curves between approximately 100 eV and above 250 eV, thus, comes mostly from the vertex corrections in W and self-consistency effects.

Two more interesting points related to the self energy figures deserve to be mentioned. The amplitude of the GW part of the self energy is about 10 times larger for the s-band and about 7 times larger for the d-band than the amplitude of the second order diagram. From this point of view, the expansion in W converges rather fast for this material, again supporting the applicability of ab-initio MBPT. Second point is that the vertex correction for the s-band is only about twice less than the correction for the d-band. Quantitatively, it is far from being negligible. Thus, for the quantitative account of the correlation effects in 3d metals one cannot consider only 3d electrons.

The effect of the vertex correction on the calculated irreducible polarizability is shown in the Fig. \ref{p_pw}, where the momentum dependence of the polarizability for zero frequency and for the first positive Matsubara frequency ($\nu=2\pi T$) is presented. Components of this function for higher frequencies have similar (to the component with $\nu=2\pi T$) behavior with gradual decreasing in the amplitude. Generally one can see a considerable effect of the vertex corrections on the calculated polarizability, reflecting the fact that total difference in the self energy calculated with scGW and scGW$\Gamma$ cannot be explained only by the second order correction to the self energy.

One more insight into the accuracy of a certain approximation when applied to a specific material is the degree of satisfaction of the Ward-Takahashi Identity (WTI).\cite{pr_78_182,inc_6_371} Particularly useful for applications in solid state physics consideration of the subject was provided in the Ref. [\onlinecite{rdnc_11_1}]. In scGW$\Gamma$ calculations, one needs only the charge vertex, whereas the current vertex also entering the generalized Ward-Takahashi Identity (see for instance the formula (7.11) in the Ref.[\onlinecite{rdnc_11_1}]) is not normally available. Thus, it is useful in practice to consider only zero external (bosonic) momentum case, when the formula (7.11) of the Ref.[\onlinecite{rdnc_11_1}] can be translated into the following identity:

\begin{equation}\label{wti1}
\Gamma^{\mathbf{k}}_{\lambda\lambda'}(\omega;\nu)=\delta_{\lambda\lambda'}+i\frac{\Sigma^{\mathbf{k}}_{\lambda\lambda'}(\omega+\nu)-\Sigma^{\mathbf{k}}_{\lambda\lambda'}(\omega)}{\nu},
\end{equation}
where $\omega$ is the fermionic Matsubara's frequency and $\nu$ is bosonic (external) Matsubara's frequency. Figure \ref{wti} represents the non-trivial part of the calculated vertex function $\Gamma$ (it is equal to zero in the GW approximation) as compared to the non-trivial part on the right hand side of the Eq. (\ref{wti1}). As one can see, the vertex correction dominates for the 6 bands closest to the Fermi level (one s-band and five d-bands) where the first order vertex correction eliminates at least 80\% of the mismatch found in scGW approximation. The remaining mismatch (about 20\%) depends very little on the full number of bands included in the vertex correction part of the calculations (if one changes this number from 10 to 20 bands). Thus, higher order diagrams not included in this study are responsible for the remaining mismatch. They can be spin-fluctuation diagrams\cite{prb_97_155128} or other diagrams and can bring some effect on the low energy band renormalization. One can speculate, however, that the effect of the higher order diagrams on the calculated equilibrium volume is very small. First reasoning for this speculation is that the calculated $V_{0}$ is very close to the experimental value and, as such, should not change much. Second reasoning is that higher order diagrams are tend to be more of atomic nature and, thus, weakly volume dependent. Yet another reasoning is that a mutual cancellation of diagrams is possible (like partial cancellation of the GW diagram and the first order vertex diagram presented in the Fig. \ref{psi_vv0}, where their volume gradient has different sign).

\section*{Conclusions}
\label{concl}

In conclusion, self-consistent vertex corrected GW calculations have been performed for vanadium metal in order to assess the feasibility of applications of ab-inito MBPT for studying GSP of realistic materials. It has been shown that scGW$\Gamma$ improves the results obtained with scGW noticeably, thus, confirming the importance of vertex corrections (despite their smallness in case of vanadium) for accurate evaluation of the ground state properties. It has been shown, that both, the correction to the irreducible polarizability and the correction to the self energy are important. Consideration of the degree of satisfaction of the Ward-Takahashi Identity leaves, however, a room for studying the higher order diagrams (such as spin-fluctuation diagrams) which could affect the calculated low energy excitations. However, the effect of higher order diagrams on the calculated GSP is, most likely, very small. From more general point of view, the work demonstrates the feasibility of using the ab-initio MBPT as one more tool (in addition to already available tools) to study the ground state properties of crystalline solids.

\section*{Acknowledgments}
\label{ackn}

This work was   supported by the U.S. Department of Energy, Office of Science, Basic
Energy Sciences as a part of the Computational Materials Science Program.

\appendix

\section{Details of the evaluation of some diagrams}\label{diag_det}

\begin{figure}[t]
\centering
\includegraphics[width=7.0 cm]{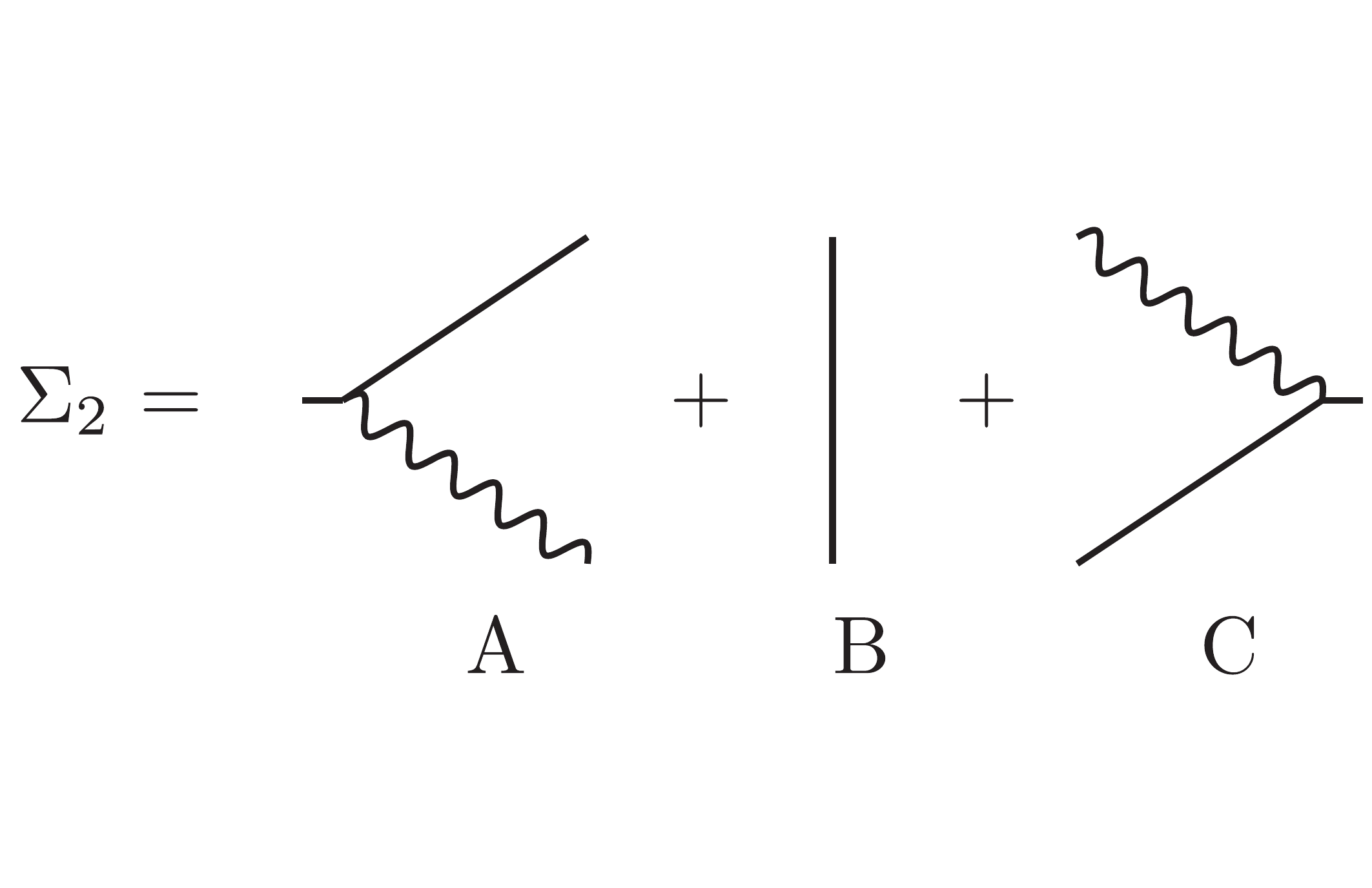}
\caption{Scheme of the evaluation of the second order diagram for the self energy.}
\label{sigma_steps}
\end{figure}

\begin{figure}[b]
\centering
\includegraphics[width=7.0 cm]{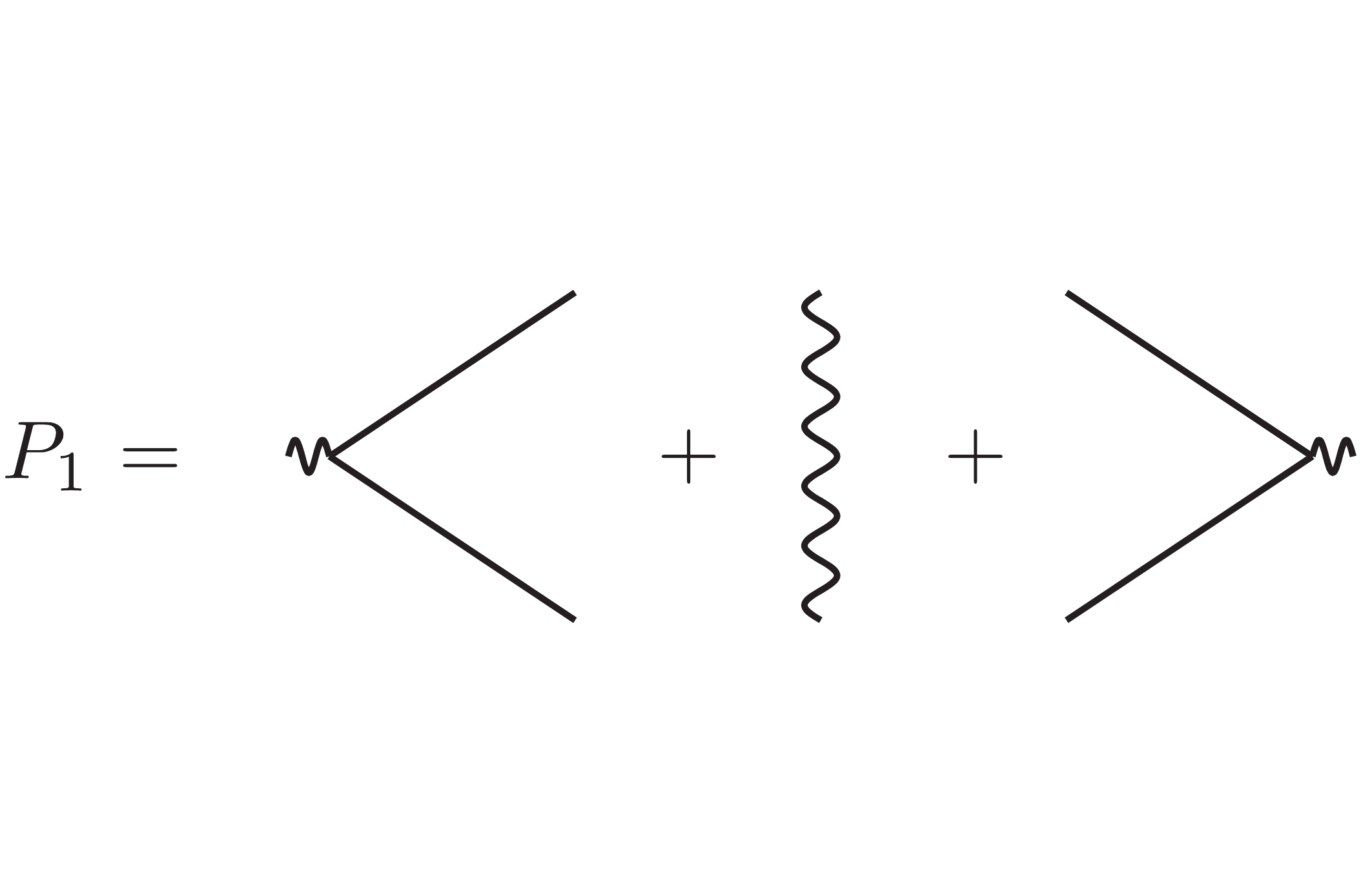}
\caption{Scheme of the evaluation of the first order diagram for the irreducible polarizability.}
\label{p_steps}
\end{figure}

In the Ref. [\onlinecite{prb_94_155101}] the evaluation of the vertex correction to the self energy was presented as a two-step process. In the first step, the non-trivial part of the three-point vertex function $\Gamma$ was evaluated. In the second step, this vertex function was combined with the Green function G and the screened interaction W to form the self energy $GW\Gamma$. However, it was found later, that a considerably more efficient procedure for the evaluation of the second order self enery consists in the evaluation of the corresponding diagram directly, avoiding the intermediate construction of $\Gamma$. Namely, the second diagram presented in Fig. \ref{diag_S} can be evaluated in three steps as it is demonstrated in Fig. \ref{sigma_steps}. The pieces A, B, and C shown in Fig. \ref{sigma_steps} are combined together beginning from the right and proceeding to the left. Essentially, the algorithm is very similar to the algorithm for the first order polarizability presented in Fig. \ref{p_steps} and described in details in Ref. [\onlinecite{prb_94_155101}].

The piece C (Fig. \ref{sigma_steps}) is evaluated in (reciprocal space + frequency) representation with the band states indexes representing the orbital basis set. After evaluation, piece C is transformed into (real space + imaginary time) representation (see Ref. [\onlinecite{prb_94_155101}] for the specifics of the representations of functions in real space). Thus, pieces B and C are combined in (real space + imaginary time) representation, which approximately can be thought of as point by point multiplication. After that, the object B+C is transformed back to the (reciprocal space + frequency) representation, and it is combined with the piece A. In practice, this algorithm of self energy evaluation is a few times faster than the original one presented in Ref. [\onlinecite{prb_94_155101}] and requires considerably less memory.


\end{document}